\documentclass[preprint2]{aastex}
\usepackage{color}
\usepackage{graphicx}% Include figure files
\usepackage{natbib}
\bibpunct{(}{)}{;}{a}{}{,}
\usepackage{hyperref} 
\newcommand{\adsurl}[1]{\href{#1}{ADS}} 
\providecommand{\url}[1]{\href{#1}{#1}} 

\slugcomment{Submitted for publication to Astrophysical Journal}

\def\beq{\begin{equation}}
\def\eeq{\end{equation}}
\def\bea{\begin{eqnarray}}
\def\eea{\end{eqnarray}}

\def\bwt{\begin{widetext}}
\def\ewt{\end{widetext}}
\def\eg{e.g.,~}

\def\black{\color{black}}
\usepackage{amssym,aas_macros}
\shorttitle{A New Proposal for a Universal Mass Function}
\shortauthors{Prescod-Weinstein \& Afshordi}

\begin{document}

\title{Using Dark Matter Haloes to Learn about Cosmic Acceleration:\\ A New Proposal for a Universal Mass Function}%What do dark matter haloes teach us about cosmic acceleration? }

\author{Chanda Prescod-Weinstein\altaffilmark{1,2}}
\affil{NASA Postdoctoral Program Fellow, Goddard Space Flight Center, 8800 Greenbelt Rd, Greenbelt, MD, 20770, USA}
\affil{Perimeter Institute for Theoretical Physics, 31 Caroline St. N, Waterloo, ON, N2L 2Y5 Canada}
\affil{Department of Physics and Astronomy, 200 University Ave. W, University of Waterloo, Waterloo, ON, N2L 3G1 Canada}

\and 

\author{Niayesh Afshordi\altaffilmark{3}}
\affil{Perimeter Institute for Theoretical Physics, 31 Caroline St. N, Waterloo, ON, N2L 2Y5 Canada}
\affil{Department of Physics and Astronomy, 200 University Ave. W, University of Waterloo, Waterloo, ON, N2L 3G1 Canada}

\altaffiltext{1}{Current address: Goddard Space Flight Center, Greenbelt, MD, 20770, USA}
\altaffiltext{2}{cosmos@post.harvard.edu}
\altaffiltext{3}{nafshordi@perimeterinstitute.ca}

%\date{\today}

\begin{abstract}
Structure formation provides a strong test of any cosmic acceleration model because a successful dark energy model must not inhibit {\black or overpredict} the development of observed large-scale structures. Traditional approaches to studies of structure formation in the presence of dark energy or a modified gravity implement a modified Press-Schechter formalism, {\black which relates the linear overdensities to the abundance of dark matter haloes {\it at the same time}.
We critically examine the universality of the Press-Schechter formalism for different cosmologies, and show that the halo abundance is best correlated with spherical linear overdensity at $94\%$ of collapse (or observation) time. We then extend this argument to ellipsoidal collapse (which decreases the fractional time of best correlation for small haloes), and show that our results agree with deviations from modified Press-Schechter formalism seen in simulated mass functions. This provides a novel universal prescription to measure linear density evolution, based on current and future observations of cluster (or dark matter) halo mass function. In particular, even observations of cluster abundance in a single epoch will constrain the entire history of linear growth of cosmological of perturbations.}
\end{abstract}

\maketitle

\section{Introduction}
\label{intro_structure}
An important part of the effort to explain cosmic acceleration and the cosmological constant problem is testing proposed models in the context of what are, at this stage, better-established physical pictures. Structure formation could prove to be an incredibly useful phenomenological method for distinguishing models of cosmic acceleration.

It is currently believed that large-scale structure formation has its seeds in small quantum fluctuations in the early universe (\eg \citet{1992PhR...215..203M}). The current model for structure formation is elegant in its fundamental simplicity. Random inhomogeneities, artifacts of cosmic inflation, create a runaway effect in which overdense regions attract more matter, thus becoming more dense and leading to galaxies, stars, and planets.

Better understanding of this process is independently an intriguing enterprize in the field of cosmology. However, in this work we focus on the relationship between the cosmic acceleration and structure formation. More specifically, different cosmological pictures (cosmologies with differing causes of acceleration, such as a cosmological constant, dark energy, or modifications of Einstein gravity) might have expansion histories that are similar to one another but leave different imprints on large-scale structures, and in particular on galaxy clusters. Therefore, structure formation provides a unique testing ground for models of cosmic acceleration (\eg \citet{2006PhRvD..74d3513I,2008PhRvD..78d3514A,2009ApJ...692.1060V}). Here we critically examine some of the assumptions in this program, and develop a framework to enhance the accuracy of this kind of work.

The first step in this direction is to revisit how the Press-Schechter formalism~\citep{1974ApJ...187..425P} (PSF) is used to predict the cluster mass function. \citet{1974ApJ...187..425P} have argued that the number density of dark matter haloes (or galaxy clusters) of mass $M$ is given by:
\beq
\frac{dn(M,z)}{dM} = f[\sigma(M,z)]\frac{\bar{\rho}_m(z)}{M}{\partial \ln\sigma^{-1}(M,z) \over \partial M},
\label{f_def}
\eeq
where $\sigma^2(M,z)$ is the variance of {\it linear} overdensity in spherical regions of mass $M$ at redshift $z$, while $\bar{\rho}_m(z)$ is the mean matter density of the universe. For random gaussian initial conditions, $f[\sigma]$ is given by:
\beq
f_{PS}[\sigma] = \sqrt{\frac{2}{\pi}}\frac{\delta_{sc}}{\sigma} \exp\left[-{\delta^2_{sc}\over 2\sigma^2}\right],
\label{eq:PS}
\eeq
where $\delta_{sc}$ ($\simeq 1.68$ for most $\Lambda$CDM cosmologies) is the spherical collapse threshold for linear overdensities \citep{1972ApJ...176....1G}.

\begin{figure}[!h]
\includegraphics[width=\linewidth]{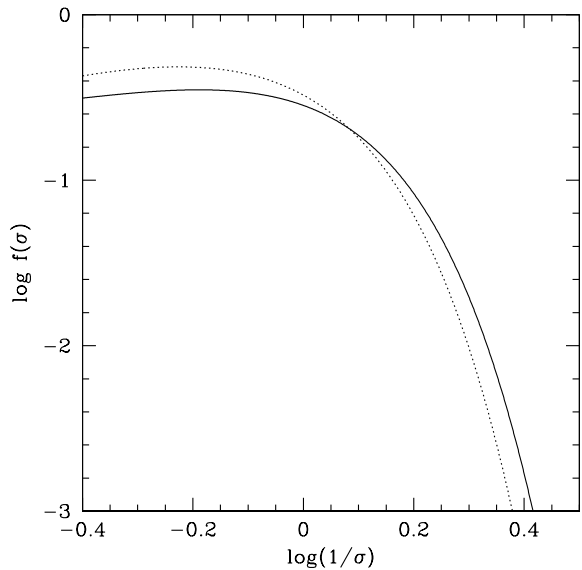}
\caption[The Press-Schechter function vs. numerical simulations]{ \footnotesize A comparison of Press-Schechter prediction for the function $f(\sigma)$ (dotted line; Eq. \ref{eq:PS}), with a parameterized fit to the numerical simulations (solid line; Eq. \ref{eq:Tinker}).}
\label{PS}
\end{figure}

While the PSF successfully predicts the broad features of the simulated cluster mass functions, it proves too simplistic for detailed model comparisons required for precision cosmology. Consequently, several authors including \citet{1999MNRAS.308..119S}, \citet{2001MNRAS.321..372J}, \citet{2002ApJ...573....7E}, \citet{2006ApJ...646..881W}, and \citet{2008ApJ...688..709T} have refined the  function $f(\sigma)$ to better match the mass functions in  N-body simulations. We shall refer to these approaches as modified PSF. For example, \citet{2006ApJ...646..881W} and \citet{2008ApJ...688..709T} propose a fitting formula of the form:
\beq
f(\sigma) = A\left[\left(\sigma\over b\right)^{-a}+1\right]e^{-c/\sigma^2},  \label{eq:Tinker}
\eeq
where $(A,a,b,c)=(0.186,1.47,2.57,1.19)$ give a good fit to simulated haloes of overdensity $\Delta=200$, in a concordance $\Lambda$CDM cosmology at $z=0$ \citep{2008ApJ...688..709T} (see Fig. \ref{PS}). While most of this work is based on fitting formulae to simulated mass functions, \citet{2001MNRAS.323....1S} argue that an approximate implementation of ellipsoidal collapse can account for most of the deviations from the PSF.

However, a more pressing question for cosmological applications is whether the  function $f(\sigma)$ is {\it universal}, or rather can vary for different cosmologies or cosmic acceleration models.
In other words, could the same modified PSF be used to accurately predict halo abundance in cosmologies with different cosmological parameters? %In other words, universality is the idea that CMF can be calculated by relying only on matter density $\Omega_{m}$ and linear density fluctuations.
While earlier studies failed to find such dependence,  \citet{2008ApJ...688..709T} first noticed a systematic evolution of $f(\sigma)$ with redshift, implying a breakdown of universality at the $10-30$\% level (also see \citet{2010arXiv1005.2239B}).
\citet{2010arXiv1001.3425C} note that universality is limited by the nonlinearity of structure formation, and the cluster mass function shows some redshift dependence at higher redshifts that can be (partially) understood in the context of spherical collapse. However, the spherical collapse model falls short of explaining the $10-30$\% deviations from universality in all but the most massive clusters (see Fig. \ref{delta_f}).

In this work, we contribute to the effort to better understand the role and limits of universality in the cluster mass function by introducing a new parameter that appears to be universal across cosmological models.\footnote{Here we define cosmological models to mean different values of the cosmological constant density $\Omega_{\Lambda}$ or different values of the dark energy equation of state parameter, $w$. For now, we do not consider quintessence models where $w$ is dynamical in nature, or possible modifications of Einstein gravity. Future work will extend our study to such possibilities.} In particular, the modified PSF relies on  $\sigma(M)$, the root-mean-square of \textit{linear} density fluctuations at the time of observation, when in reality, observed clusters are very non-linear objects with overdensities exceeding $200$. We thus seek to find a universal time in the past when we could make a connection between the nonlinear structures that we observe in the present and the linear structures that existed in the past, since all structures go through a linear phase. Our basic strategy is to find the time in the past at which the linear density of the structures that collapse today show minimum dispersion, as we vary cosmologies.

In \S~\ref{linear}, we introduce linear perturbation theory for modeling structure formation which leads to a linear differential equation.

In \S~\ref{nonlinear}, we present the complete nonlinear differential equation that governs the growth of matter perturbations in spherical overdense regions in the presence of a cosmological constant.

In \S~\ref{computation}, we describe a numerical method which we developed in order to solve both the nonlinear and linear structure formation equations in the presence of a cosmological constant, {\black or dark energy with a constant equation of state}.

In \S~\ref{lambdaresults}, we discuss the {\black implications of our numerical study, i.e. we find a universal fraction of collapse time, where linear density is the same independent of cosmology.}
%We then seek to generalize from a cosmological constant to models of dynamical dark energy. \S~\ref{varyingw} presents the results of this study.

In \S~\ref{cmf} we reexamine the idea of universality of mass functions in light of the results of the previous section, including the effect of ellipsoidal collapse on the formalism, and propose a new mass function for general dark energy models.

Finally, in \S~\ref{strucconclusions}, we conclude with an overview of our results and a discussion of future prospects.

%----------------------------------------------------------------------
\section{Background: Linear Perturbations}
\label{linear}
%----------------------------------------------------------------------
%We assume the spherical top-hat as our collapse model. In this approach, small scales see collapse first, leading to the production of larger scale structures due to mergers of the smaller objects. An alternative approach via the Zeld'ovich approximation is not used because it was found (REFERENCE) to be less consistent with the cold dark matter model. While in the spherical model, small scales collapse first, in the Zeldovich approximation, small structures are by-products of the formation of larger ones. This is more consistent with hot dark matter, a dark matter model that has essentially been dismissed.

The linear perturbation theory that is used to describe structure formation can be derived via a fluid picture. We use a Newtonian treatment because when density perturbations are small, the gravitational potential will be nonrelativistic \citep{1993ppc..book.....P}. The standard equations of fluid dynamics are reviewed here. First, the continuity equation:
\begin{equation}
\label{linear1}
\frac{\partial \rho}{\partial t} + \nabla \cdot (\rho \vec{v})=0,
\end{equation}
where $\rho$ is the matter density and $\vec{v}$ is the fluid velocity.
The Euler equation is:
\begin{equation}
\label{linear2}
\frac{\partial \vec{v}}{\partial t} + (\vec{v} \cdot \nabla)\vec{v} = -\frac{\nabla P}{\rho} - \nabla \phi,
\end{equation}
where $\phi$ is the gravitational potential.
Finally, the Poisson equation is:
\begin{equation}
\label{linear3}
\nabla^{2}\phi=4\pi G_{N}\rho.
\end{equation}
%Using the Hubble flow equation {\blue in an expanding universe}:
%\begin{equation}
%\label{linear4}
%\vec{v}(\vec{r},t)=H(t)\vec{r},
%\end{equation}
Introducing the comoving coordinates $\vec{x}=\frac{\vec{r}}{a(t)}$, we can write the velocity in terms of the comoving coordinates and the scale factor $a(t)$:
\begin{equation}
\label{linear5}
\vec{v}=\frac{d\vec{r}}{dt}=\dot{a}\vec{x}+\dot{\vec{x}}a=\frac{\dot{a}}{a}\vec{r}+\vec{u}(\frac{\vec{r}}{a},t)
\end{equation}
where $\vec{u}=\dot{\vec{x}}a(t)$ is the peculiar velocity. %Note that we want a partial time derivative that respects comoving coordinates, i.e. one that keeps $\vec{x}$ fixed:
%\begin{equation}
%\label{linear6}
%\left(\frac{\partial \rho(\vec{r},t)}{\partial t}\right)_{r}=\left(\frac{\partial {\rho(\frac{\vec{r}}{a},t)}}{\partial t}\right)_{r}.
%\end{equation}
Note that at constant $\vec{r}$, $\frac{d\vec{r}}{dt}=0$ so $\dot{a}\vec{x}+a\dot{\vec{x}}=0$, giving us $\frac{\dot{a}}{a}\vec{x}=-\frac{\partial\vec{x}}{\partial t}$.
This leads us to the following relation:
\bea
\label{linear7}
\left(\frac{\partial \rho(\vec{r},t)}{\partial t}\right)_{r}=\left(\frac{\partial \rho(\vec{r},t)}{\partial t}\right)_{x}+\frac{\partial \vec{x}}{\partial t}\cdot\frac{\partial \rho}{\partial \vec{x}}\nonumber\\=\left(\frac{\partial \rho(\vec{r},t)}{\partial t}\right)_{x}-\frac{\dot{a}}{a}\vec{x}\cdot\nabla_{x}\rho(\vec{x},t).
\eea

We now wish to rewrite Equation~\ref{linear1} in terms of the comoving coordinates, which essentially means replacing the partial differential with the modified one from Equation~\ref{linear7}. Moreover, what we are really interested in is the development of relative deviations from the mean density, or the density contrast: $\delta=\frac{\rho}{\overline{\rho}}-1$. Thus, we write $\rho = \overline{\rho}(1+\delta)$ and assuming that $\overline{\rho}$ is {\black the mean density of regular matter}, we expect $\overline{\rho}\propto a^{-3}$. This gives:
\begin{equation}
\label{linear8}
0=\left(\frac{\partial \delta}{\partial t}\right)_{\vec{x}}+\frac{1}{a}\nabla\cdot{(1+\delta)\vec{u}}.
\end{equation}
We make similar transformations for the Poisson and Euler equations. Next, we drop higher order terms (e.g. $O(u^{2})$ or $u\delta$). We differentiate the linearized continuity equation and take the divergence of the linearized Euler equation. This gives us a differential equation that depends entirely on $\delta$ and not on {\black $\vec{u}$}:
\begin{equation}
\ddot{\delta}+2\frac{\dot{a}}{a}\dot{\delta}=4\pi G_{N} \overline{\rho}\delta .
\label{eq:linear_delta}
\end{equation}
The linear growth factor $D(t)$  is defined as the growing solution for $\delta$ in this equation.

%----------------------------------------------------------------------
\section{$\Lambda$ \& Non-linear Structure Formation}
\label{nonlinear}
%----------------------------------------------------------------------
In \S~\ref{linear}, we derived the differential equation that governs the growth of linear matter perturbations. We used a familiar fluid dynamics picture to do so. Here we derive the full non-linear equation for spherical overdensities using only cosmological considerations. It should be noted that this particular form of the non-linear equation is only strictly valid for $\Lambda$CDM cosmologies, where the inside of a spherical top-hat overdensity can be considered as a separate universe. More complex models such as dark energy models with different values of $w$ require additional considerations which will be {\black discussed below}.

We consider a physical picture in which structure formation arises due to a uniformly positive spherical perturbation away from an average matter density, i.e. a top-hat matter overdensity. This scenario is similar to considering two cosmologies with two distinct scale factors: one for the outer universe and another for the overdense region. Of course, we are interested in a scenario where a dark energy component similar to a cosmological constant is at play, so we will assume the presence of one as part of our base model.

For the external universe, we write the Friedmann equation with zero curvature:
\begin{equation}
\label{nonlinear1}
\left(\frac{\dot{a}_{o}}{a_{o}} \right)^{2}=\frac{8\pi G_{N}}{3}(\overline{\rho}+\rho_{\Lambda})=H^{2}.
\end{equation}
As we did in \S~\ref{linear}, we can assume $\overline{\rho}\propto a^{-3}$ for ordinary matter, while $\rho_{\Lambda} = $ const. denotes the cosmological constant density. We also note that the value of the cosmological constant will be the same inside the overdense region and the background.

In a general scenario with dynamical dark energy models we cannot assume that curvature, often denoted by $k$, will be a constant inside the overdense region due to the presence of pressure gradients. Therefore, \citet{1998ApJ...508..483W} point out that we are compelled to use the time-time component of Einstein's equations, as these do not explicitly involve the curvature term. However, in the presence of a cosmological constant, or $w=-1$, we can ignore these considerations and begin with the first order Friedmann equation. We then calculate $k$, which can be seen as an integration constant that arises in going from second to first order Friedmann formulations, using initial conditions.

The scale factor in the overdense region is governed by:
\begin{equation}
\label{nonlinear2}
\left(\frac{\dot{a}_{i}}{a_{i}}\right)^{2} + \frac{k}{a_{i}^{2}}=\frac{8\pi G_{N}}{3}(\rho_{i}+\rho_{\Lambda}).
\end{equation}
Again, we define $\rho_{i} = \overline{\rho}(1+\delta)$. A little bit of algebra yields the following full differential equation for $\delta$:
\begin{equation}
\label{nonlinear3}
\left[\overline{\rho}(1+\delta)\right]^{-\frac{2}{3}}\left[-\frac{8\pi G_{N}}{3}\delta \overline{\rho} - \frac{2}{3}\frac{H\dot{\delta}}{1+\delta} + \frac{1}{9}\frac{\dot{\delta}^{2}}{(1+\delta)^{2}}\right]+k=0
\end{equation}

It is important to reiterate the importance of having access to both linear and nonlinear solutions. As noted by \citet{2010MNRAS.tmp.1011P} amongst others, although initially both the complete solution and its linear approximation will track, eventually the nonlinear solution will grow much faster relative to the scale factor.

Following \citet{2000cils.book.....L}, we can show that knowing nonlinear theory is necessary. A critical point in the evolution of a structure's collapse is the turnaround event in which universal expansion's dominance over the perturbation is eclipsed by gravitational collapse. In other words, at the turnaround point, a potential structure has detached from background expansion, but complete gravitationally-bound structure formation has not yet begun. This might be thought as the true birth of a structure within the void. Knowing the nonlinear solution allows us to find out the value of the scale factor and the overdensity at this so-called ``turn-around point.''

\begin{figure}[!h]
\includegraphics[width=\linewidth]{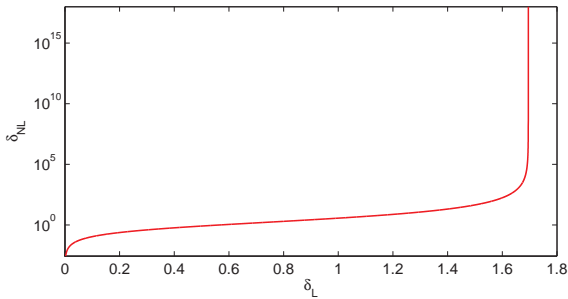}
\caption[Spherical collapse in the Einstein-de Sitter Universe]
{\footnotesize This plot shows nonlinear vs. linear overdensities in the Einstein-de Sitter Universe. The code produces, as expected, a $\delta_{NL}$ that diverges when $\delta_{L} \approx 1.68$. }
\label{eds}
\end{figure}

Numerically, at the point of complete collapse the nonlinear solution will ``blow up'' and approach infinity {\black (see Fig. \ref{eds})}. The linear density at this point, $\delta_{sc}$ is the quantity that enters the original Press-Schechter formalism (see \S~\ref{intro_structure}) and is used as proxy between linear and non-linear structures. Physically, we do not expect an actual singularity. This ``blow up'' point is considered to be the beginning of virialization, a process whereby energy in the bulk infall of matter is redistributed into random motion of dark matter particles, leading to a system in virial equilibrium, where kinetic energy $T$ and the potential energy $U$ are related by the virial theorem:
\begin{equation}
\label{nonlinear4}
T_{vir}=\frac{1}{2}(R ~\partial U/\partial R)_{vir},
\end{equation}
(see \citet{2005JCAP...07..003M}). For the purposes of this study, more details on this process are not necessary.

We can generalize to cases where $w\neq-1$, i.e. non-cosmological constant scenarios. \citet{2007JCAP...11..012A} provides a comprehensive derivation of the full non-linear equation, which we refer the interested reader to for complete details. For our purposes, it will suffice to show the final result, which is Equation 7 in \citet{2007JCAP...11..012A}:
\bea
\label{dynamicalw1}
\ddot{\delta}_{j}+\left(2H-\frac{\dot{w}_{j}}{1+w_{j}} \right)\dot{\delta}_{j} -\frac{4+3w_{j}}{3(1+w_{j})}\frac{\dot{\delta_{j}}^{2}}{1+\delta_{j}}\nonumber\\-4\pi G \sum_{k}{\overline{\rho}_{k}(1+w_{k})(1+3w_{k})\delta_{k} (1+\delta_{k})} = 0.
\eea
The subscripts $j$ and $k$ refer to different fluids in the system, e.g. matter and dark energy. In a scenario where the dark energy can clump, this becomes a system of equations. As we discuss below, we do not allow this possibility. Therefore, noting that $w=0$ for  matter, we get the following non-linear equation governing the behavior of matter perturbations:
\begin{equation}
\ddot{\delta}_{m}+2H\dot{\delta}_{m} -\frac{4}{3}\frac{\dot{\delta}_{m}^{2}}{1+\delta_{m}}-4\pi G\overline{\rho}_{m}\delta_{m} (1+\delta_{m}) = 0.
\label{eq:nonlinear_delta}
\end{equation}
But, why exactly are we allowed to ignore the clumping in dark energy?

The discussion about dark energy perturbations is often cast in terms of the effective speed of sound for dark energy, \eg \citet{2004PhRvD..69h3503B}.
%A subtlety here involves understanding how to calculate the effective speed of sound for the dark energy fluid, which is an important characteristic for determining clustering properties.
Typically one may expect that for an adiabatic fluid with a constant equation state parameter $w$, the speed of sound is given by $c^{2}=\delta P/\delta\rho = w$. However, when $w_{de}<0$, clearly this becomes imaginary, suggesting a catastrophic instability, \eg see \citet{2005PhRvD..72f5024A}, and thus we must use a more general definition of the $c$. In other words, dark energy with constant $w$ cannot be an adiabatic fluid, implying $w= \frac{P}{\rho} \neq \frac{\delta p}{\delta \rho}$. While the equation of state parameter can still give us information about the background evolution, the full action of the fluid is necessary to compute its effective speed of sound: $c_{eff}^{2}\equiv \frac{\delta p}{\delta \rho}$. It turns out that for the simplest quintessence models $c_{eff}=1$, although for more general actions $c_{eff}$ could essentially take any value, \eg \citet{2001PhRvD..63j3510A}.

As to the question of dark energy clumping, we know that pressure fluctuations propagate with the speed of sound $c_{eff}$. Therefore, dark energy should be smooth on scales smaller than its sound horizon $\sim c_{eff}/H$. As long as $c_{eff} \sim 1$, all the collapsed structures at late times are much smaller than the sound horizon, implying that dark energy perturbations $\delta_{DE}$ should be negligible for their formation.

%----------------------------------------------------------------------
\section{Numerical Techniques}
\label{computation}
%----------------------------------------------------------------------

Although \citet{1972ApJ...176....1G} showed that the perturbation equation with the assumption of spherical symmetry (also known as the spherical top-hat problem) can be solved analytically for the case of an Einstein-de Sitter (EdS) universe ($\Lambda = 0$), we are interested in cases where the cosmological constant/dark energy are non-zero. Therefore, a numerical solution is necessary.

We built a code in C++ that utilizes the Runge-Kutta method for numerical solutions of ordinary differential equations. The full code can be found in {\black \citet{chandesis}}. The code runs an instance of a loop for value of $\Omega_{\Lambda,{present}}$ in the range of 0 to 0.7 (the currently measured value of dark energy density), which solves the second-order linear differential equation, Equation~\ref{eq:linear_delta}, as well as the full non-linear equation, Equation~\ref{nonlinear3}. In order to solve both equations, the solution to the differential equation for the background scale factor, Equation~\ref{nonlinear1}, is found. The curvature constant is calculated {\black for the initial overdensity} before proceeding to a solution for both the linear and nonlinear differential equations. %We set the present Hubble parameter to one. In other words, all time steps are in units of the present value of the Hubble time.

Collapse time is formally defined to be the time at which nonlinear overdensity, $\delta_{NL}$, goes to infinity. However, we define collapse numerically by requiring a {\black large value of nonlinear overdensity, $\delta_{NL} =200$}. The initial conditions are calibrated such that they provide the same results as the analytical EdS model, namely that $\delta_{L}(t_{collapse})\approx 1.7$. This is essentially done by assuming that at early times $\delta$ scales linearly with the scale factor $a$, as expected for the growing solution to Equation (\ref{eq:linear_delta}) in the matter era \footnote{It is important to remember that $\delta$ is a ratio of two numbers whose units are that of density. Therefore, $\delta$ is dimensionless.}.

%Recording the value of $\delta_{linear}$, the scale factor, and the time at this point, we then rescale such that time is given as a ratio with the collapse time, %the scale factor as a ratio with the collapse scale factor and $\delta_{L}$ as a ratio with its value at collapse.

%----------------------------------------------------------------------
\section{Results}
\label{lambdaresults}
%----------------------------------------------------------------------
As stated in the Introduction, we wish to discover the time (as a fraction of the collapse time)  such that the variance of $\delta_{L}(t)$ is at a minimum, as we vary cosmologies. In doing so, we find the time in cosmological history where linear theory is most likely to accurately predict gravitational collapse, independent of cosmology.

For simplicity, we chose to plot the relative difference of $\delta_L(t)$ for different cosmologies with respect to the fraction of time to the collapse time in the Einstein-de Sitter universe. However, it should be noted that the results are independent of this choice, and one could easily calibrate with respect to a universe with a non-zero $\Lambda$ instead. Having found a common ground, all the data was searched for a single point in time (in units of collapse time) where the variance of $\frac{\delta_{L}(Q)}{\delta_{L}(\Lambda=0)}-1$ was at a minimum, where $Q$ stands for different dark energy models under consideration, whether a general fluid or a cosmological constant.

\begin{figure}[!h]
\includegraphics[width=\linewidth]{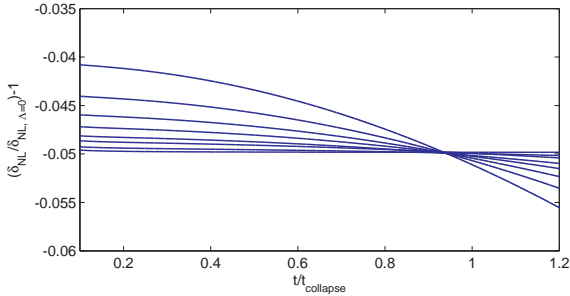}
\caption[Comparison of collapses in different $\Lambda$ cosmologies]
{\footnotesize Relative change in $\delta_{L}(t/t_{collapse})$ in $\Lambda$CDM cosmologies ($\Omega_{\Lambda} =0.1,0.2,...,0.7$), with respect to the Einstein-de Sitter Universe. $t_{collapse}$ is calculated for spherical overdensities. The curves seem to intersect at $t/t_{collapse} \simeq 0.94$, and a calculation of the point of minimum variance between the lines confirms this.}
\label{eds2}
\end{figure}

The interpolations and variance computations were carried out using a script in MatLab, which can be found in \citet{chandesis}. In the case of the simple cosmological constant models, Fig.~\ref{eds2} shows the result that at $\frac{t}{t_{collapse}}\simeq 0.94$, the variance in $\delta_L(t/t_{collapse})$ for different cosmological constant models hits a minimum of $1.80 \times 10^{-9}$. Values of $\delta_{L}$ at range from 1.602 for Einstein-de Sitter \footnote{This is below the predicted analytic value of 1.68 at collapse because while $\delta_{L}=1.68$ is expected as $\delta_{NL}\rightarrow \infty$, we have set the collapse to occur at $\delta_{NL}=200$, resulting in a lowered collapse $\delta_L$.} to 1.599 in the $\Lambda=0.7$ universe.

Computing the solution to Equation (\ref{eq:nonlinear_delta}) requires some modifications to the code. Instead of scanning over different cosmological constant values, this version of the code varies between constant values of $w$. Moreover, as current observations, \eg \citet{2010arXiv1001.4538K} set a (very conservative) upper limit of $-1/3$ for the value of $w$ , we studied cases where $w$ was smaller than $-1/3$.

Because the cosmological constant is a special case of this scenario with $w=-1$, we are able to check the self-consistency of our methods (Eq. \ref{nonlinear3} vs. Eq. \ref{eq:nonlinear_delta}) finding that both versions of the code produce the same results for a universe with $\Omega_{\Lambda} = 0.7$ (approximately the universe that we live in).

\begin{figure}[!h]
\includegraphics[width=\linewidth]{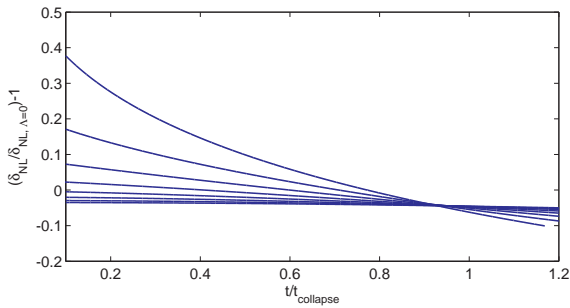}
\caption[Comparison of collapses in different $w$ cosmologies]
{\footnotesize Same as Fig. (\ref{eds2}), but with varying values of dark energy equation of state ($w=-0.3,-0.4,...,-1$), at $\Omega_{\Lambda}(today)=0.7$. Again, curves seem to intersect at $t/t_{collapse} \simeq 0.94$, and, again, a calculation of the point of minimum variance between the lines confirms this. We begin at the value -0.3 because current data constrains the parameter to be smaller than this value.}
\label{eds3}
\end{figure}

Fig. (\ref{eds3}) shows a similar comparison to that of Fig. (\ref{eds2}), but with different values of equation of state, $w$, with $\Omega_{DE} =0.7$. Interestingly, we can again clearly see a point of minimum variance at $t/t_{collapse} \simeq 0.94$, suggesting that this result might be quite independent of the dark energy model (at least within the spherical collapse approximation). The variance at this minimum is $7.5 \times 10^{-7}$. In the next section, we discuss the physical significance of this result.

%----------------------------------------------------------------------
\section{What does the Cluster Mass Function teach us about cosmology?}
\label{cmf}
%----------------------------------------------------------------------
At first, one might be puzzled by the fact that $\delta_L$ happens to have almost the same value at 94\% of the collapse time, independent of $w$ or $\Lambda$, even though the linear approximation breaks down long before this point. In other words, why should non-linear collapse show such strong correlation with the linear evolution? This can be understood as the near cancelation of two different effects with opposite signs:
\begin{enumerate}
\item With the exact same initial conditions, the presence of dark energy weakens the gravitational attraction near the turn around point, which in turn stretches the collapse time.
\item The linear growth factor $D(t)$, which is the growing solution to Equation (\ref{eq:linear_delta}), slows down as dark energy starts to dominate, since the Hubble friction $2H\delta$ stops decaying ($H \rightarrow$ const.), while matter density decays exponentially $\bar{\rho} \propto a^{-3} \propto \exp(-3Ht)$.
\end{enumerate}
It turns out that for near $\Lambda$CDM cosmologies, these two effects nearly cancel each other, i.e. the linear growth is slowed down, but $t_{collapse}$ is also longer, resulting in nearly the same value of $\delta_L$ close to (i.e. at 94\% of) the collapse time.

\begin{figure}[!h]
\includegraphics[width=\linewidth]{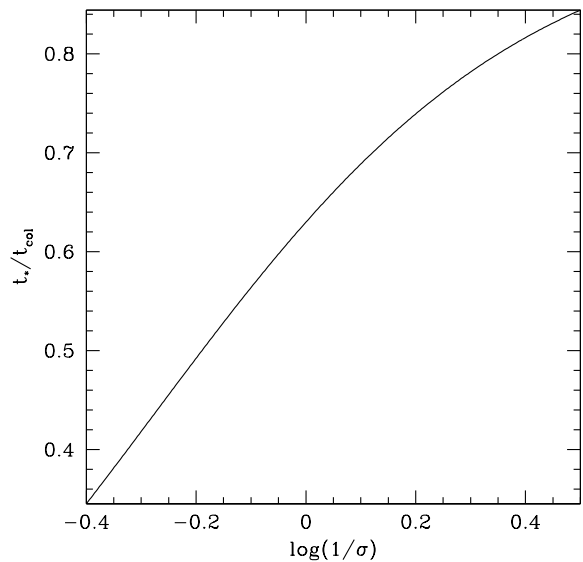}
\caption[Time of minimum variance in linear overdensity]{\footnotesize The expected time of minimum variance in linear overdensity, in units of ellipsoidal collapse time. Observing the cluster number counts at $t_{col}$ should tell us the linear overdensity of the collapsing region at $t_*$, independent of the dark energy model. }
\label{tcol}
\end{figure}

However, we should note that this result is specific to the spherical collapse scenario. On the other hand, real collapsing regions are far from spherical. Even though very rare peaks of a random gaussian field can be approximated as spheres, more abundant haloes could significantly deviate from sphericity, \eg \citet{1986ApJ...304...15B}. For Einstein-de Sitter universe, \citet{2001MNRAS.323....1S} give a simple numerical fit for the impact of ellipticity on the linear collapse threshold, $\delta_{ec}$:
\beq
\delta_{ec} \simeq \delta_{sc}\left[1+ \sigma(M)^{1.23}\right],
\eeq
where $\delta_{sc}\simeq 1.686$ is the spherical collapse threshold. Since $\delta_L \propto t^{2/3}$ in an Einstein-de Sitter universe, this implies that, for the same initial overdensity, the collapse time of an elliptical region is longer than that of a spherical region by a factor of:
\beq
\frac{t_{collapse, elliptical}}{t_{collapse, spherical}} \simeq  \left[1+ \sigma(M)^{1.23}\right]^{3/2}.\label{t_ratio}
\eeq
In other words, we need to extrapolate the linear theory predictions in Figs. (\ref{eds2}-\ref{eds3}) farther beyond the point of intersection to actually hit gravitational collapse. Therefore, {\black assuming that Equation (\ref{t_ratio}) is approximately independent of the dark energy model, it can be combined with} the results of previous sections to show that the point of minimum variance in $\delta_L$ is shifted to smaller values of $t/t_{collapse}$, i.e.:
\beq
\frac{t_*}{t_{collapse}} = \frac{0.94}{\left[1+ \sigma(M)^{1.23}\right]^{3/2}},\label{eq:t*}
\eeq
if we include the impact of ellipsoidal collapse. This result is shown in Fig. (\ref{tcol}), and demonstrates how measuring the mass function of clusters at a given era may tell us about the entire history of linear growth, and not just a snapshot at the time of observation (as is in the traditional universal mass function hypothesis).

Inspired by the fitting formula Eq. (\ref{eq:Tinker}), Equation (\ref{eq:t*}) leads us to propose a new universal cluster mass function:
\beq
f(\sigma;Q) \simeq g(\sigma) e^{-h(\sigma)/\sigma^2_*}, \label{new_f}
\sigma_*=\sigma D\left[0.94\times  t_o\over (1+\sigma^{1.23})^{3/2}\right],
\eeq
where the actual mass function is related to $f(\sigma;Q)$ through equation (\ref{f_def}), and $t_o$ is the time of observation at which $D(t)$ is normalized to unity.
In other words, Equation (\ref{new_f}) suggests that the exponential cut-off in the cluster number counts at any time $t$ is set by the linear density fluctuations $\sigma_*$ at an earlier time $t_*$, which is set by Equation (\ref{eq:t*}). As suggested by several recent numerical studies (see Introduction) $f(\sigma;Q)$ depends on cosmology (denoted by $Q$), but we propose the functions $g$ and $h$ to be universal, while the dependence on cosmology (or dark energy models) only enters through $D(t_*)$.
The explicit dependence of $g$ and $h$ on $\sigma$ at the time of observation is justified, as the value of $\sigma$ acts as a proxy for the asphericity of the collapsing region~\citep{2001MNRAS.323....1S}.

\begin{figure}[!h]
\includegraphics[width=\linewidth]{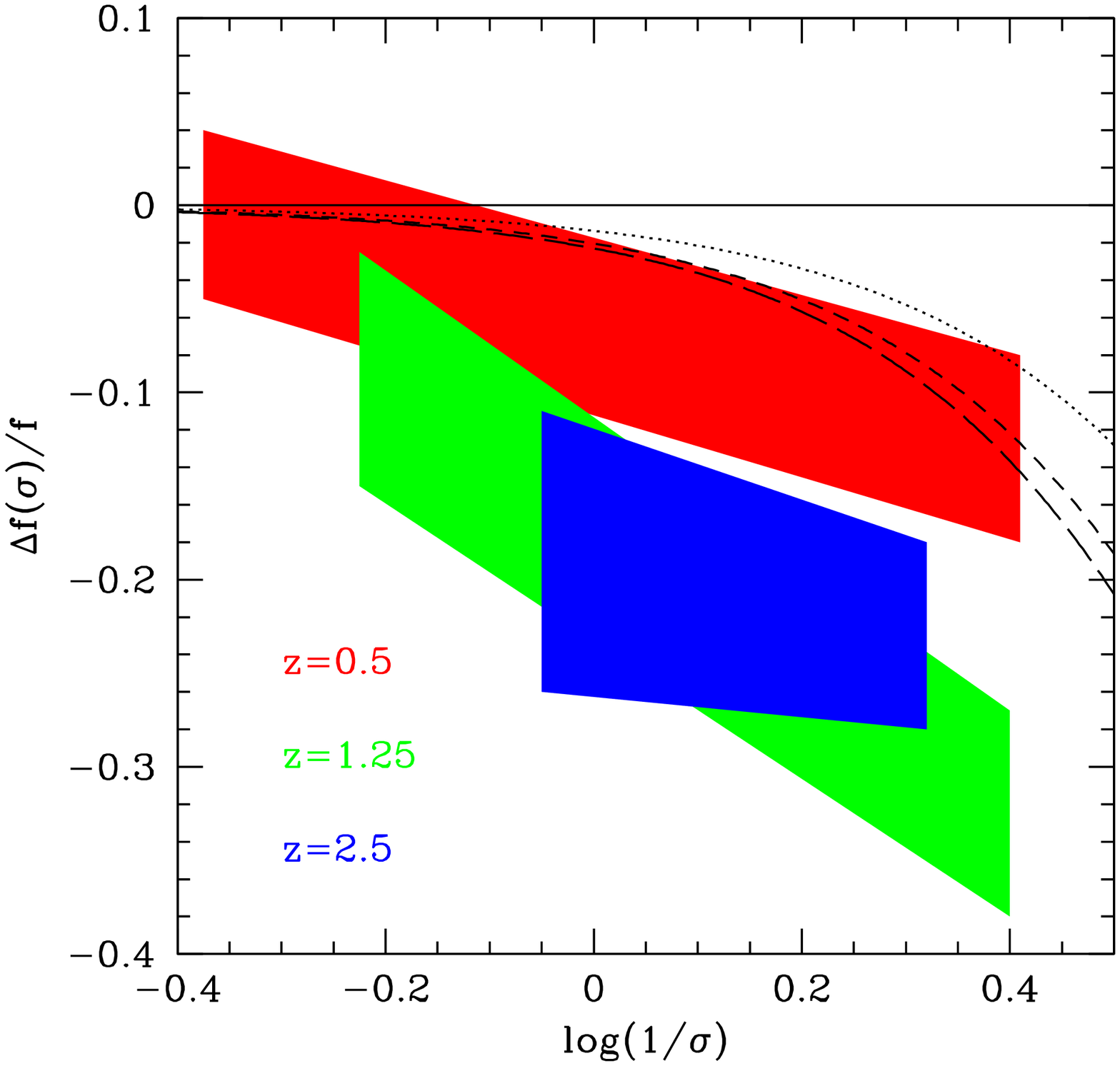}
\includegraphics[width=\linewidth]{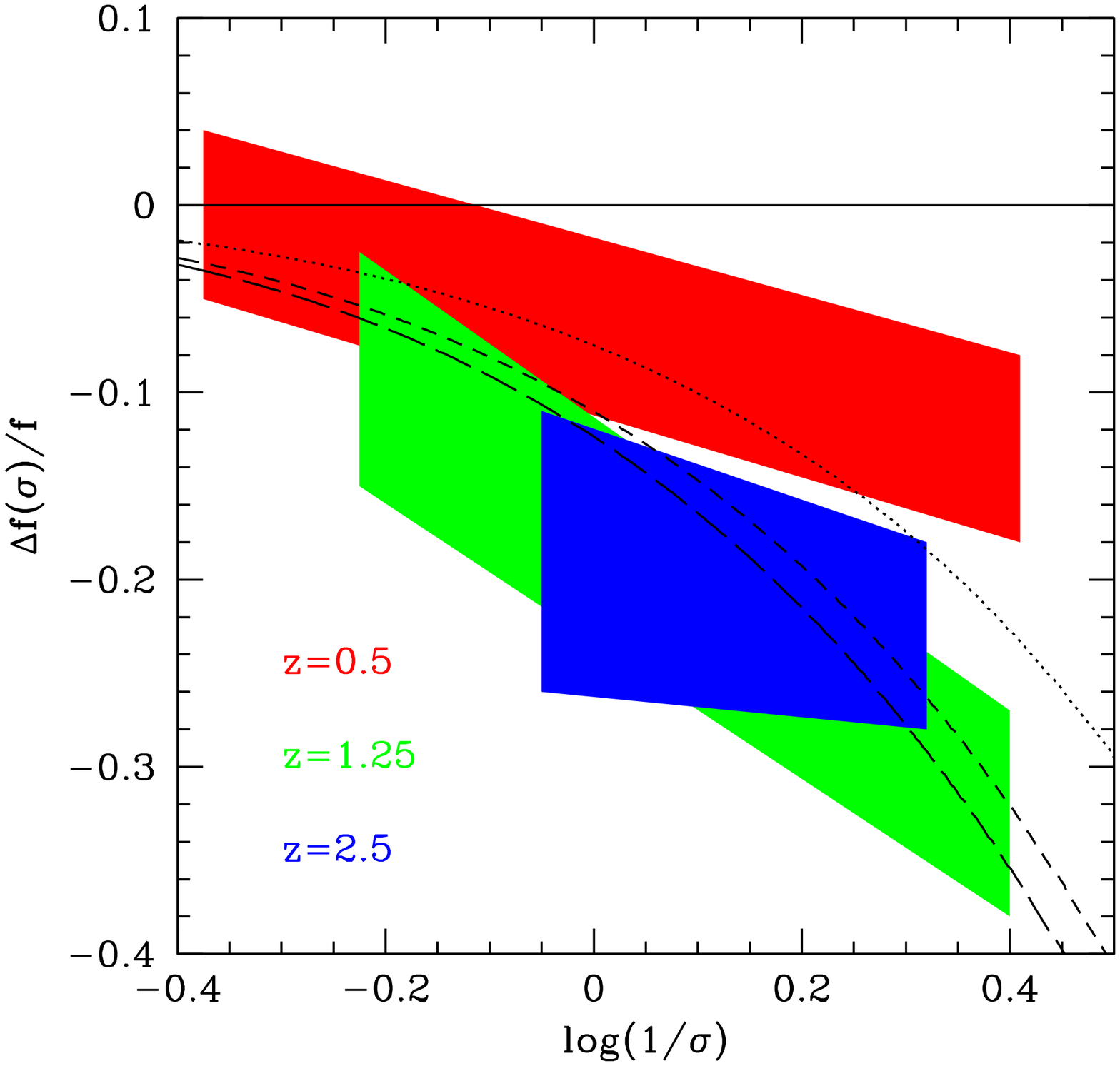}
\caption[Relative change in $f(\sigma)$ at high redshifts]{\footnotesize The relative change in $f(\sigma)$ at high redshifts, compared to $z=0$. The colored regions show the simulated results from \citet{2008ApJ...688..709T}. Curves in top panel show our analytic prediction without ellipticity corrections, while curves in the bottom panel include the ellipticity corrections (Eq. \ref{new_f}). The solid, dotted, short-dashed, and long-dashed curves refer to $z=$ 0, 0.5, 1.25, and 2.5 respectively. }
\label{delta_f}
\end{figure}

Comparing to Eq. (\ref{eq:Tinker}), we fix $g$ and $h$ as:
\beq
g(\sigma)= A\left[\left(\sigma\over b\right)^{-a}+1\right], h(\sigma)= c D^2\left[0.94\times  t_o\over (1+\sigma^{1.23})^{3/2}\right]_{\Omega_{\Lambda}=0.7},\label{hsigma}
\eeq
where $(A,a,b,c)=(0.186,1.47,2.57,1.19)$  \footnote{We recognize that these choices for $g$ and $h$ are not unique, but as Fig. (\ref{delta_f}) demonstrates, these are sufficient to fit current simulations}.

Fig. (\ref{delta_f}) compares the evolution in the function $f(\sigma;Q)$ in N-body simulations, with our prediction from Eqs. (\ref{new_f}-\ref{hsigma}).  We see that while the result from spherical collapse, $t_*=0.94 \times t_{collpase}$, underpredicts the evolution (top panel), Eq. (\ref{new_f}), which includes the ellipsoidal correction to the collapse time (Eq. \ref{eq:t*}), can successfully reproduce the evolution in $f(\sigma;Q)$ (bottom panel).

%----------------------------------------------------------------------
\section{Conclusions and Future Prospects}
\label{strucconclusions}
%----------------------------------------------------------------------
We have presented a study of non-linear gravitational collapse in different models of dark energy/cosmic acceleration. In particular, we critically examined the correlation between the linear growth of fluctuations and the emergence and statistics of collapsed objects (such as dark matter haloes or galaxy clusters). First, we focused on the collapse of spherical overdensities, and discovered that they all have the same linear ovderdensity ($\simeq 1.50$), at $\simeq 94\%$ of the  time of collapse/virialization, independent of the density or equation of state of dark energy. We then used a simple prescription to include the impact of ellipsoidal collapse in our finding, and then used this result to propose a new universal mass function for galaxy clusters/dark matter haloes. Our semi-analytic predictions are consistent with the observed evolution in mass function of haloes in N-body simulations.

Future work will include the adaptation of this prescription to study models with dynamical equations of state, such as quintessence or modified gravity models. A particularly challenging application would be to the gravitational aether/black hole model \citep{2009PhRvD..80d3513P}. Because of the way dark energy is sourced  in the gravitational aether model, there are unique {\black technical} challenges associated with properly describing structure formation in its presence.

Finally, we recognize that a more systematic approach to the question of universal mass functions should be possible, and given the level of theoretical and observational activities in this field, it is unlikely that the present work provides the last word on this subject. However, the novel (albeit trivial) observation of this work is that measuring the cluster mass function will teach us about the entire history of linear growth. This is in contrast to many previous cosmological applications of cluster mass functions, which assume a one-to-one correspondence with the linear $\sigma(M)$ at the time of observation. Similar to a multi-level archeological excavation, dark matter haloes can be thought of as artifacts of linear growth. As Fig. (\ref{tcol}) and Eq. (\ref{new_f}) suggest, the number of smaller haloes (with larger $\sigma(M)$) can teach us about the early evolution of linear growth, while the bigger haloes (with smaller $\sigma(M)$) tell us about its more recent history.

We thus speculate that this perspective can eventually lead to yet untapped information about the nature and history of cosmic acceleration, especially as the releases of several large scale cluster surveys such as Atacama Cosmology Telescope (ACT) \citep{2010arXiv1006.5126M}, South Pole Telescope (SPT) \citep{2010arXiv1006.3068A}, Planck \citep{2007MNRAS.382..158G}, and Red Sequence Cluster Sequence 2 (RCS2) \citep{2007astro.ph..1839Y} are now looming on the horizon.

\subsection*{Acknowledgements}
We would like to acknowledge helpful comments and discussions with Robert Brandenberger, Gil Holder, Brian McNamara and Ryan Morris. During the time that this work was conducted, CP and NA were supported by Perimeter Institute for Theoretical Physics. Research at Perimeter Institute is supported by the Government of Canada through Industry Canada and by the Province of Ontario through the Ministry of Research \& Innovation. CP is supported by an appointment to the NASA Postdoctoral Program at the Goddard Space Flight Center, administered by Oak Ridge Associated Universities through a contract with NASA.

\bibliography{universal_mass}

\end{document}